\documentclass[aps,twocolumn,superscriptaddress]{revtex4-1}
\usepackage{pdfpages}
\usepackage{amsmath,amssymb,graphicx,braket,hyperref}

\newcommand{\bk}{{\bf k}}
\newcommand{\psiL}{\psi^{\rm L}}
\newcommand{\psiR}{\psi^{\rm R}}

\makeatletter
\AtBeginDocument{\let\LS@rot\@undefined}
\makeatother

\begin{document}

\title{Topological Band Theory for Non-Hermitian Hamiltonians}

\author{Huitao Shen}
\affiliation{Department of Physics, Massachusetts Institute of Technology, Cambridge, Massachusetts 02139, USA}
\author{Bo Zhen}
\affiliation{Department of Physics, Massachusetts Institute of Technology, Cambridge, Massachusetts 02139, USA}
\affiliation{Department of Physics and Astronomy, University of Pennsylvania, Philadelphia, Pennsylvania 19104, USA}
\author{Liang Fu}
\affiliation{Department of Physics, Massachusetts Institute of Technology, Cambridge, Massachusetts 02139, USA}

\begin{abstract}
We develop the topological band theory for systems described by non-Hermitian Hamiltonians, whose energy spectra are generally complex. After generalizing the notion of gapped band structures to the non-Hermitian case, we classify ``gapped'' bands in one and two dimensions by explicitly finding their topological invariants. We find nontrivial generalizations of the Chern number in two dimensions, and a new classification in one dimension, whose topology is determined by the energy dispersion rather than the energy eigenstates. 
We then study the bulk-edge correspondence and the topological phase transition in two dimensions. Different from the Hermitian case, the transition generically involves an extended intermediate phase with complex-energy band degeneracies at isolated ``exceptional points'' in momentum space. We also systematically classify all types of band degeneracies. 
\end{abstract}


\maketitle
Topological band theory provides a unified framework for a wide range of topological states of quantum matter \cite{Thouless1982,Kane2005,Moore2007, Fu2007a, Fu2007, Schnyder2009, Kitaev2009a, Bansil2016, Chiu2016, Armitage2017} such as insulators, (semi)metals and superconductors, and of classical wave systems \cite{Lu2013,Lu2016,Kane2013,Susstrunk2015} such as photonic crystals and mechanical metamaterials. In this theory, band structures of periodic media are classified by topological invariants associated with energy eigenstates in the momentum space. A well-known example is the TKNN invariant or Chern number \cite{Thouless1982,Kohmoto1985} for band structures in two dimensions with an energy gap. An important consequence of this classification is that the interface between topologically inequivalent media necessarily hosts gapless boundary states, whereby the topological invariant changes its value.

Studies of topological band theory have so far mostly dealt with systems described by Hermitian Hamiltonians. Recently there has been a growing interest in topological properties of non-Hermitian Hamiltonians \cite{Esaki2011,Liang2013,Lee2016,Leykam2017,Menke2017,Xu2017,Gonzalez2017,Hu2017,Xiong2017} applicable to a wide range of systems such as (but not limited to) systems with open boundaries
\cite{Rotter2009,Choi2010,Zhen2015,Cao2015a,Gao2015,San-Jose2016}
and systems with gain and/or loss
\cite{Bender1998,Klaiman2008,Makris2008,Longhi2009,Guo2009,Ruter2010,Lin2011a,Liertzer2012,Feng2012,Regensburger2012,Peng2014,Chang2014,Brandstetter2014,Hodaei2014,Peng2014a,Feng2014,Menke2017}.
Interestingly, non-Hermitian systems have unique topological properties with no Hermitian counterparts.
A fascinating example is non-Hermitian Hamiltonians at exceptional points, where two or more eigenstates coalesce \cite{Keldysh1971,Kato1966,Moiseyev2011,Berry2004,Rotter2009,Heiss2012}. Very recently, the topological nature of exceptional points in non-Hermitian Hamiltonians with additional symmetries have been recognized \cite{Esaki2011,Liang2013,Lee2016,Leykam2017,Menke2017}. Dynamical phenomena near exceptional points are also being explored both theoretically \cite{Uzdin2011,Berry2011,Berry2011a,Gilary2013,Graefe2013,Kapralova-Zdanska2014,Milburn2015,Hassan2017} and experimentally \cite{Doppler2016,Xu2016a}.

In this work, we develop the topological band theory for non-Hermitian Hamiltonians and explore its consequences, highlighting unique features due to non-Hermiticity. We start by defining the notion of ``gapped'' non-Hermitian band structures whose energy spectrum is generally complex. We then classify topologically distinct ``gapped'' band structures and topologically stable band degeneracies.
Non-Hermitian bands with nonzero Chern numbers in two dimensions are shown to support protected edge states, with a range of energies connecting two bulk bands in the complex plane. A new topological invariant unique to non-Hermitian band structures is found from the energy dispersion, instead of Bloch wavefunctions. Furthermore, we find that the topological phase transition between distinct ``gapped'' non-Hermitian Hamiltonians generally involves an intermediate phase with band degeneracies at isolated points in momentum space, leading to the first realization of exceptional points in two-dimensional band structures.

Consider a non-Hermitian Hamiltonian of a periodic system, whose eigenstates are Bloch waves and whose energies $E_n(\bk)$ vary with crystal momentum $ \bk $ in the Brillouin zone (BZ), thus defining a band structure. Here $ n $ is the band index that labels different eigenstates.
While $E_n(\bk)$ are generally complex,
we define a band $ n $ to be ``separable'' if its energy $E_n(\bk) \neq E_m(\bk)$ for all $m \neq n$ and all $\bk$.
We define a band $ n $ to be ``isolated'' if $E_n(\bk) \neq E_m(\bk')$ for all $m \neq n$ and all $\bk, \bk'$, i.e., the region of energies $\{ E_n(\bk), \bk \in {\rm BZ} \}$ in the complex plane does not overlap with that of any other band.
In this case, we say the band $E_n(\bk)$ is surrounded by a ``gap'' in the complex energy plane where no bulk states exist. 
A band is called ``inseparable'' if at some momentum the complex-energy is degenerate with another band. 
Our definition of  ``separable'', ``isolated'' and ``inseparable'' bands are mathematically natural generalizations of the gapped, fully gapped and gapless bands in the Hermitian case, and form the basis of our topological classification to be presented below.

\textit{Chern Numbers in 2D Separable Bands}
Associated with each separable band is a set of energy eigenstates defined over the BZ.
Topological invariants, such as the (first) Chern number for an energy band in two dimensions, can be constructed from these eigenstates in a similar way as in Hermitian systems.

However, an important difference now is the left eigenstate and right eigenstate of a non-Hermitian matrix $H \neq H^\dagger$ are generally unrelated, although they share the same eigenvalue. The right and left eigenstates satisfy the following eigenvalue equations:
\begin{equation}
H\ket{\psiR_n} = E_n\ket{\psiR_n}, \;
H^\dagger \ket{\psiL_n}  = E_n^* \ket{\psiL_n}
\end{equation}
respectively. For separable band structures, one can prove that $ \braket{\psiL_n|\psiR_n}\neq 0 $ (Supplemental Material Sec.~I). Thus for any separable band with energy $ E_n $ in two dimensions $ \bk\equiv(k_x,k_y) $, one can construct four different gauge invariant Berry curvatures:
\begin{equation}
B^{\alpha\beta}_{n,ij}(\bk)\equiv i\braket{\partial_i \psi^{\alpha}_n(\bk)|\partial_j \psi^{\beta}_n(\bk) },
\end{equation}
with the normalization condition $ \braket{ \psi^{\alpha}_n|\psi^{\beta}_n}=1 $. $ \alpha,\beta ={\rm L}/{\rm R} $. We refer to $ B^{\rm LL}, B^{\rm LR}, B^{\rm RL} $ and $ B^{\rm RR} $ as ``left-left'', ``left-right'', ``right-left'' and ``right-right'' Berry curvatures.  

The integrals of these four Berry curvatures over the BZ define four seemingly different Chern numbers:
\begin{equation}
N^{\alpha\beta}_n=\frac{1}{2\pi}\int_{\rm BZ} \epsilon_{ij}B^{\alpha\beta}_{n,ij}(\bk)d^2\bk,
\label{eq:cdeff}
\end{equation}
where $ \epsilon_{ij}=-\epsilon_{ji} $. Importantly, we prove all four Chern numbers are equal $ N^{\rm LL}=N^{\rm LR}=N^{\rm RL}=N^{\rm RR} $, implying that the topology is captured by a single Chern number.
We emphasize that these four Berry curvatures are indeed locally different quantities, although their integrals all yield the same Chern number.
The proof is presented in Supplemental Material Sec.~II. These Chern numbers will vanish if $ H(\bk)=H(\bk)^{\rm T} $ or $H(\bk)=H(-\bk)^{\rm T}  $ (Supplemental Material Sec.~III). 

A remarkable universal result of the topological band theory in Hermitian systems is the existence of topologically protected edge states localized at the interface between two topologically distinct gapped phases, with energies inside the band gap. For non-Hermitian Hamiltonians, we ask whether topological edge states exist, and if so, what are their energies in the \textit{complex} plane.

For concreteness, we first show the existence of topological edge states in a generalized two-dimensional Dirac fermion model with non-Hermitian terms:
\begin{equation}
H(\bk) = (k_x+i \kappa_x)\sigma_x + (k_y+i \kappa_y)\sigma_y + (m+i\delta)\sigma_z,
\label{eq:Ham2D}
\end{equation}
The energy dispersion of $H$ is obtained by diagonalization:
\begin{eqnarray}
E_\pm(\bk)=\pm\sqrt{k^2 - \kappa^2 +m^2 - \delta^2+2i( \bk \cdot {\boldsymbol\kappa}+ m\delta)}, \nonumber
\label{ek}
\end{eqnarray}
with $k \equiv |\bk|$, $ {\boldsymbol \kappa} \equiv (\kappa_x, \kappa_y) $ and $\kappa \equiv |\boldsymbol \kappa|$.
For $ \kappa <|m|$, this complex-energy band structure is separable by our definition above. It then follows from continuity that the separable bands at $m< - \kappa$ and $m > \kappa$ are adiabatically connected to the gapped bands in the Hermitian limit $\delta=\kappa= 0$ with $m<0$ and $m>0$ respectively by tuning $ \kappa $ to zero, and therefore are topologically distinct with Chern numbers differing by $1$.

To demonstrate the existence of topological edge states, we solve the domain wall problem, where two semi-infinite domains with different parameters $ (\boldsymbol{\kappa}_1, m_1, \delta_1)  $ and $ (\boldsymbol{\kappa}_2, m_2, \delta_2)  $ are separated by a domain wall along the $ y $ axis.
Since the momentum parallel to the interface $ k_y$ is conserved, we can write the edge state wavefunction as $\psi_{k_y}(x, y) = e^{i k_{y}y } \psi_{k_{y}}(x)$ and
solve the one-dimensional generalized Dirac equation for $\psi_{k_{y}}(x)$:
\begin{widetext}
\begin{equation}
[(-i\partial_x+i\kappa_x(x) )\sigma_x+(k_y+i\kappa_y(x) )\sigma_y+(m(x)+i\delta(x) )\sigma_z] \psi_{k_y}(x) = E_{k_y} \psi_{k_y}(x),
\label{eq:domain}
\end{equation}
\end{widetext}
where the parameters $ (\boldsymbol{\kappa}(x), m(x), \delta(x))=(\boldsymbol{\kappa}_1, m_1, \delta_1)\theta(-x)+(\boldsymbol{\kappa}_2, m_2, \delta_2)\theta(x) $ take respective values in the regions $x>0$ and $x<0$. $ \theta(x) $ is the step function. $\psi_{k_y}(x)$ is required to be continuous at the interface $x=0$.

The solution of Eq.~\eqref{eq:domain} with the step-like domain wall takes the following form
\begin{equation}
\psi_{k_y}(x) = \begin{pmatrix}
\psi_1 \\ \psi_2
\end{pmatrix}
\left[\exp(x/\lambda_+)\theta(-x)+\exp(x/\lambda_-)\theta(x)\right].
\label{eq:ansatz}
\end{equation}
Localized edge states only exist when $\mathrm{Re}(1/\lambda_+) >0$ and $\mathrm{Re}(1/\lambda_-) <0$.

Solving $ \lambda_{\pm} $ for the most general case is complicated. For $ \kappa_y=0 $, we can obtain the analytical solution when the Dirac mass $ m $ have opposite signs in the regions $ x<0 $ and $ x>0 $. The localization lengths are \footnote{We note that the edge state exists even in the inseparable phase when $ \kappa_{1,x}<-|m_1| $ or $ \kappa_{2,x}>|m_2| $. These states are defective in the sense that the corresponding left eigenstates do not exist. This finding is in accordance with the results in Ref.~\cite{Leykam2017}.}
\begin{equation}
\begin{split}
1/\lambda_+=&|m_1|+\kappa_{1,x}+is_1\delta_1, \\
1/\lambda_-=&-|m_2|+\kappa_{2,x}-is_2\delta_2.
\end{split}
\label{eq:sol}
\end{equation}
Here $ s_i=m_i/|m_i| $ is the sign of the Dirac mass. The dispersion of these edge state is still $ E_{k_y}= s_2 k_y$ as in the Hermitian case. Comparing Eq.~\eqref{eq:sol} with the solution in the Hermitian limit, a nonzero $ \kappa_{x,i} $ modifies the edge state localization length.  The requirements on the sign of $\mathrm{Re}(1/\lambda_\pm) $ are satisfied only for separable band structures $|\kappa_{x,i}|<|m_i|$. 

\begin{figure}[tbp]
\includegraphics[width=0.95\columnwidth]{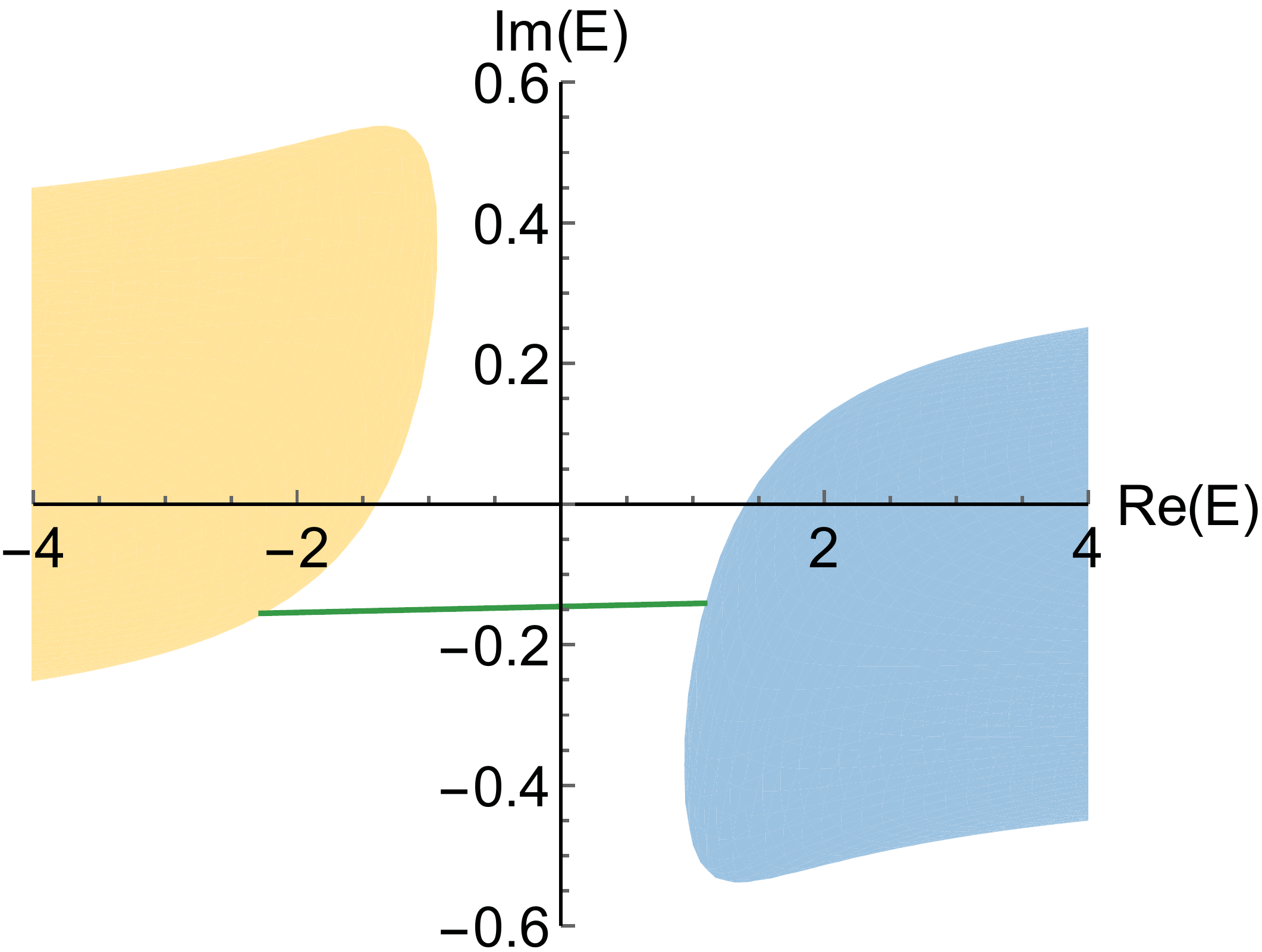}
\caption{(Color online) The energies of two bulk bands (yellow and blue regions), and the edge state (green line) in the complex-energy plane for the domain wall problem Eq.~\eqref{eq:domain}. The bulk band is isolated according to our definition. The energy unit is $ m $. The bulk phase $ \boldsymbol{\kappa}=(0.2,0.3), \delta=0.4 $ is connected to the vacuum (dispersion not shown) $ m_{\mathrm{vac}}/m=-1, \boldsymbol{\kappa}_{\mathrm{vac}}=\delta_{\mathrm{vac}}=0 $. }
\label{fig:connect}
\end{figure}

For general cases $\kappa_x, \kappa_y, \delta \neq 0$, we find numerically that when the two domains have topologically distinct separable band structures, there exists a band of edge states localized at the domain wall. The energies of these edge states have both real and imaginary parts, which lie inside the ``gap'' in the complex energy plane and connect to bulk bands. Fig.~\ref{fig:connect} shows an example of the complex-energy spectra for bulk and topological edge states in our domain wall setup. A detailed discussion on the numerics, along with the discussion of a similar lattice model, can be found in Supplemental Material Sec.~IV.

\textit{Vorticity of Energy Eigenvalues}
In addition to the Chern number, we find a new topological invariant associated with the energy dispersion of non-Hermitian band structures, rather than the energy eigenstates. Enabled by complex rather than real energies, this invariant $\nu_{mn}(\Gamma)$ is defined for any pair of the bands as the winding number of their energies $E_m(\bk)$ and $E_n(\bk)$ in the complex energy plane :
\begin{equation}
\nu_{mn}(\Gamma)=-\frac{1}{2\pi}\oint_{\Gamma} \nabla_\bk \arg\left[E_m (\bk)-E_n(\bk)\right] \cdot d \bk,
\label{eq:nu}
\end{equation}
where $\Gamma$ is a closed loop in momentum space. We call $\nu_{mn}(\Gamma)$ the vorticity. In the following, the subscript is suppressed when the band indices $ m $ and $ n $ are evident. 

A nonzero vorticity defined on a contractible loop $\Gamma$ in the BZ implies the existence of a band degeneracy within the region enclosed by $\Gamma$, where $E_m(\bk_0) = E_n(\bk_0)$.
For a pair of separable bands, the vorticity can be nonzero only for non-contractible loops in the BZ. As we will see, this leads to a $ (\mathbf{Z}/2)^d $ classification of $ d $-dimensional separable bands.
For example, consider the non-Hermitian Hamiltonian in one dimension
\begin{equation}
H(k)=b_+(k)\sigma^+ +b_-(k)\sigma^-,
\label{eq:Ham1d}
\end{equation}
where $\sigma^\pm \equiv \sigma_x \pm i\sigma_y$ and $b_{\pm}(k)$ are complex functions of $k$ with periodicity $ 2\pi $.
The spectrum of $H(k)$ is
\begin{equation}
E_\pm(k) = \pm 2\sqrt{b_+(k) b_-(k)}.
\label{eq:disp1d}
\end{equation}
The two bands are separable when $b_\pm(k) \neq 0$ for $k \in [0,2\pi]$. Taking $\Gamma$ to be the entire one-dimensional BZ, the vorticity $ \nu_\Gamma $ is simply half the sum of winding numbers of $b_+(k)$ and $b_-(k)$ around the origin of the complex plane. Although the winding of $ b_+(k) $ and $ b_-(k) $ are always integers due to periodicity, the vorticity $ \nu_\Gamma $ can be a half-integer, and is quantized as $ \mathbf{Z}/2 $.

\begin{figure}[tbp]
\includegraphics[width=\columnwidth]{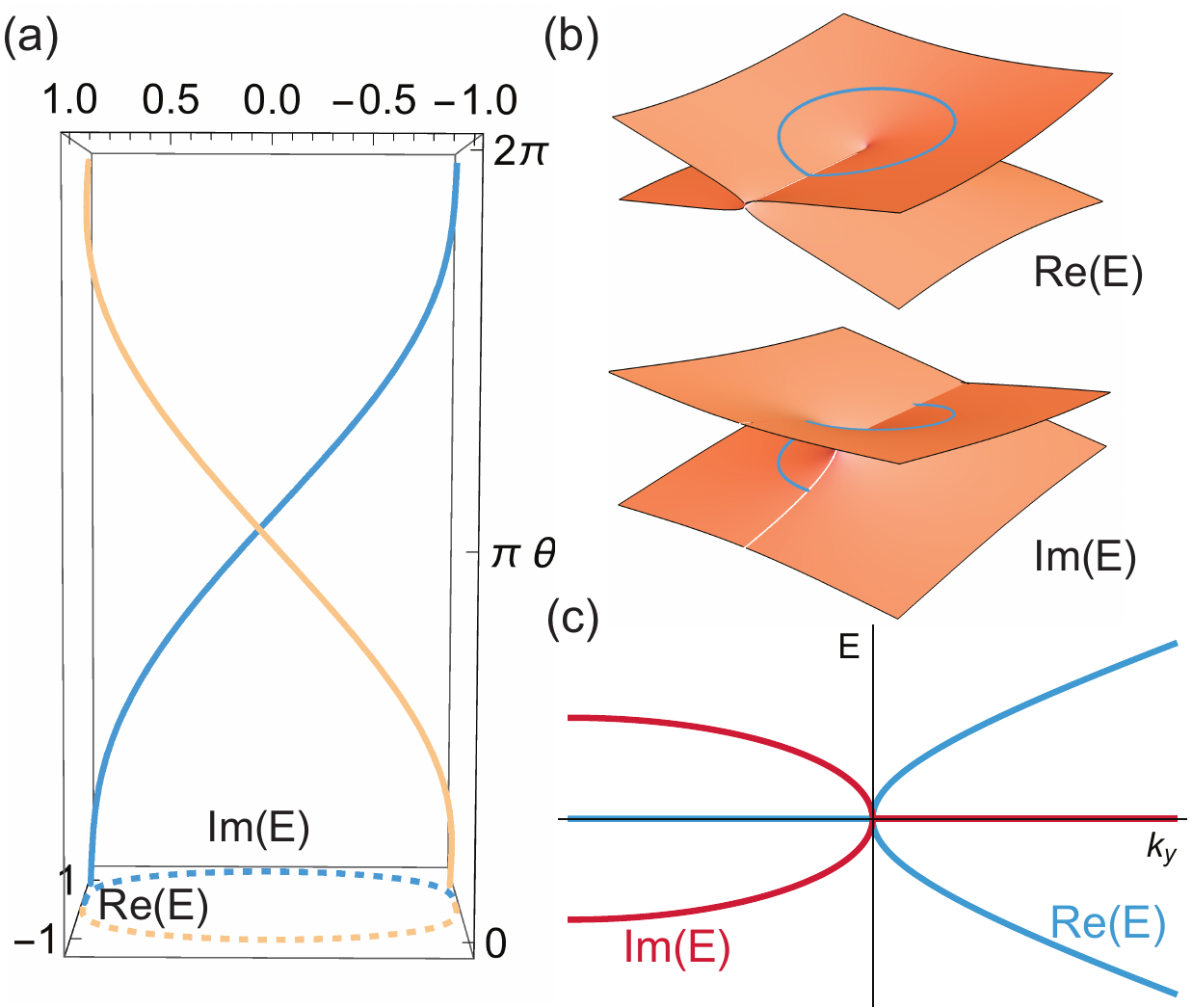}
\caption{(Color online) (a) The swapping of energy eigenvalues. $ \theta\in[0,2\pi] $ parametrizes the loop $ \Gamma $. The dashed curves are the projection of the energy trajectory. (b) The dispersion near an exceptional point. The Hamiltonian is $ H(\bk)= \sigma_+ + (k_x \sigma_x +  k_y \sigma_y) $. The loop $ \Gamma $ in (a) is the circle $ k=\sqrt{k_x^2+k_y^2}=1 $, which is parametrized by $ \theta $ as $ \bk=(\cos\theta,\sin\theta) $. (c) The energy dispersion along $ k_x=0 $. }
\label{fig:ep}
\end{figure}

It is important to notice the square root singularity in the dispersion of Eq.~\eqref{eq:disp1d}. Due to this singularity, when $ \nu_\Gamma $ is a half-integer, both the pair of energy eigenvalues $(E_+, E_-)$ and the corresponding eigenstates $(\ket{\psi_+}, \ket{\psi_-})$ are swapped without encountering any degeneracy as the momentum is traversed along $ \Gamma $ \cite{Dembowski2001, Dembowski2004}. Fig.~\ref{fig:ep}(a) shows such a scenario of $ \nu_\Gamma=1/2 $. 

The $\mathbf{Z}/2$ classification we found for separable non-Hermitian Hamiltonians in one dimension is in contrast with the case of gapped Hermitian Hamiltonians, all of which are topologically trivial.

In one dimension, there is no topologically protected edge state within the ``gap'' in the complex energy plane. Without chiral symmetry, one can always add on-site potential to lift the energy of the edge state into the bulk spectrum. We note that the zero modes found in \cite{Lee2016,Leykam2017} are due to the chiral symmetry, and our understanding is in accordance with \cite{Xiong2017}.

\textit{Topologically Stable Band Degeneracies}
Having completed the classification of separable band structures, we now study topologically stable band degeneracies in non-Hermitian systems, which cannot be removed by small perturbations. In Hermitian systems, a famous example of topologically stable band degeneracies is the Weyl point in three dimensions \cite{Armitage2017}, whereas band degeneracies in two dimensions such as the Dirac point are unstable in the absence of symmetry.  The stability of Weyl point is intimately related to the fact that finding a level degeneracy in a Hermitian matrix generically requires tuning $3$ parameters. Since energy eigenvalues of non-Hermitian Hamiltonians are complex, one might expect finding a level degeneracy requires tuning even more parameters. Remarkably, the contrary is true. For non-Hermitian Hamiltonians, finding a level degeneracy generically requires tuning $2$ parameters \cite{Berry2004}. Also, the Hamiltonian at the generic degeneracy points are defective, i.e., its entire set of eigenstates do not span the full Hilbert space. A pedagogical review of these results is in Supplemental Material Sec.~V. 

Therefore, non-Hermitian periodic Hamiltonians in two or higher dimensions can have a new type of stable band degeneracy at defective points, which has no analog in Hermitian band structures.
The $k\cdot p$ Hamiltonian near such a defective point takes the following standard form, up to a unitary transformation,
\begin{equation}
H (\bk)= a I + \epsilon \sigma_+ +\sum_{i,j}k_ic_{ij}\sigma_j,
\label{eq:Hep}
\end{equation}
where $ i=x,y $, $ j=x,y,z $,  $a$, $ \epsilon$ and $ c_{ij} $ are complex numbers.
The dispersion to the leading order of $ \bk $ is
\begin{equation}
E_\pm(\bk)= a \pm\sqrt{c_x k_x+ c_yk_y}, 
\label{eq:disp}
\end{equation}
where $ c_x=2 \epsilon (c_{xx}+ic_{xy}) $ and $ c_y=2 \epsilon (c_{yx}+ic_{yy})$. The degeneracy is defective if $ \epsilon\neq 0 $.
In the general case $c_x, c_y \neq 0$ and ${\rm Im} (c_y/c_x) \neq 0$, the band degeneracy defined by Eq.~\eqref{eq:Hep} and \eqref{eq:disp} is called an ``exceptional point'' in the literature \cite{Kato1966,Moiseyev2011,Berry2004,Rotter2009,Heiss2012}. A concrete example of a $k\cdot p$ Hamiltonian near an exceptional point is $ H(\bk)= \epsilon \sigma_+ + v(k_x \sigma_x +  k_y \sigma_y) $, whose dispersion is shown in Fig.~\ref{fig:ep}(b). 

Contrary to their name of ``exceptional'', we find exceptional points to be ubiquitous in non-Hermitian band structures in dimensions greater than one.
In particular, exceptional points appear in topological phase transitions in two dimensions, giving rise to a inseparable intermediate phase. Hermitian Hamiltonians in two dimensions do not have robust band degeneracies in the absence of symmetry.

\begin{figure}[tbp]
\includegraphics[width=1.05\columnwidth]{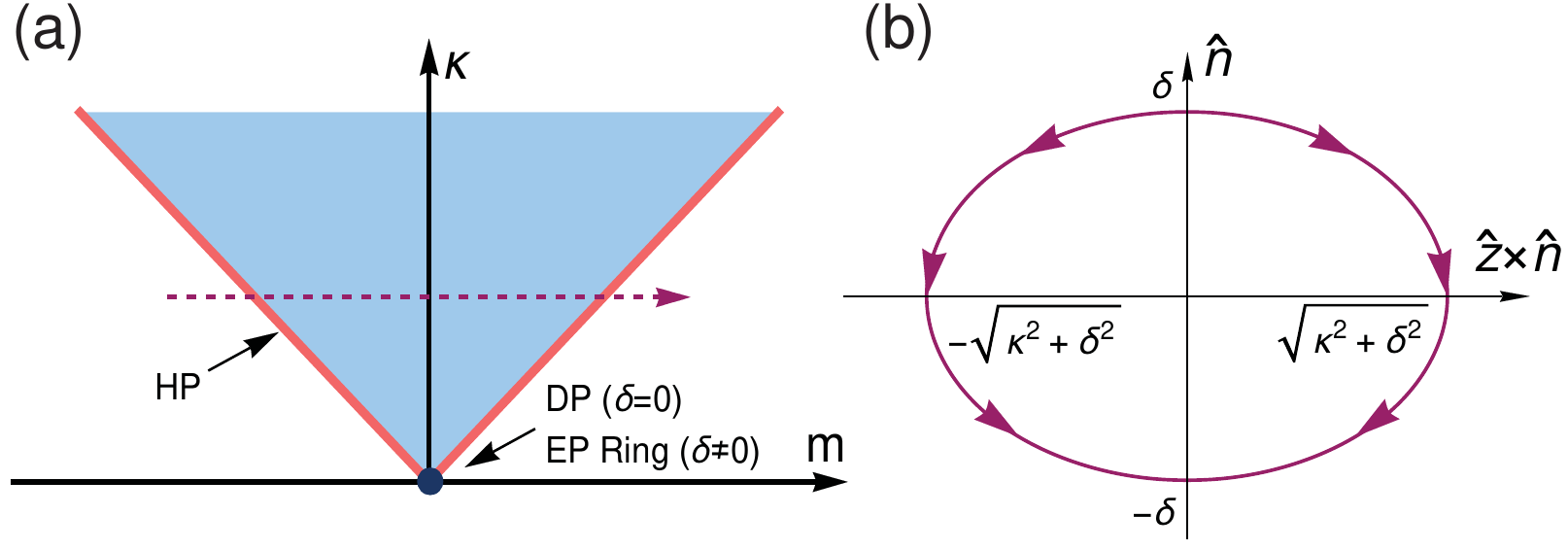}
\caption{(Color online) (a) The bulk phase diagram of Hamiltonian Eq.~\eqref{eq:Ham2D} for a given $ \delta $. The white regions represent the separable phases, and the colored region represents the inseparable phase. The light blue region $ \kappa>|m|>0 $ is the phase with a pair of exceptional points (EP Pair); the red lines $ \kappa=|m|>0 $ is the phase with a hybrid point (HP). The origin $ \kappa=m=0 $ is a Dirac point (DP) if $ \delta=0 $ and is a ring of exceptional points (EP Ring) if $ \delta\neq0 $.
(b) The trajectory of the exceptional points in the momentum space when $ m $ moves along the purple dashed line in (a). Here $ \boldsymbol{\kappa}=(\delta,0 )$. }
\label{fig:phase2D}
\end{figure}

Our claim can be demonstrated using the generalized Dirac model Eq.~\eqref{eq:Ham2D}.
The intermediate regime $ |m| < |\kappa|$ separates the two topologically distinct separable band structures at $ m>\kappa $ and $ m<-\kappa $. In this intermediate regime, the two bands $E_\pm(\bk)$ cross at two isolated points $ \bk_\pm $ in the momentum space:
\begin{equation}
\bk_\pm=-\frac{m\delta}{\kappa}\hat{\mathbf{n}}\pm \frac{\sqrt{(\kappa^2-m^2)(\kappa^2+\delta^2)}}{\kappa} \hat{\mathbf{z}} \times \hat{\mathbf{n}}.
\end{equation}
Here $ \hat{\mathbf{n}}\equiv \boldsymbol{\kappa}/\kappa $. It is straightforward to check that $\bk_\pm  $ are exceptional points. Generated from a separable band structure with zero total vorticity, these two exceptional points have opposite vorticities. The phase diagram of Eq.~\eqref{eq:Ham2D} and the typical trajectory of these two band degeneracy points are shown in Fig.~\ref{fig:phase2D}.

When $ \kappa=|m|\neq0 $ the exceptional point pair inevitably merges at
\begin{equation}
\mathbf{Q}_s=\bk_\pm=-s\delta \hat{\mathbf{n}},
\end{equation}
where $ s=m/|m| $. Denote $\mathbf{q} = \bk - \mathbf{Q}_s$. The dispersion near such a degeneracy reads
\begin{equation}
E_{s,\pm}(\mathbf{q})=\pm\sqrt{(q^2+2sq_n\delta)+2i q_nm}.
\end{equation}
$ q_n\equiv \mathbf{q}\cdot \hat{\mathbf{n}}$ is the component of $ \mathbf{q} $ along $ \hat{\mathbf{n}} $ direction. The Hamiltonian is defective at this degeneracy. However, it belongs to the case  $ c_y= {\rm Im} (c_y/c_x) = 0 $ in Eq.~\eqref{eq:disp}. The dispersion is proportional to $ \sqrt{q} $ and $ q $ along the direction of $ \hat{\mathbf{n}} $ and $ \hat{\mathbf{z}} \times \hat{\mathbf{n}} $, resulting in a zero vorticity. Being defective but with no vorticity distinguishes this degeneracy from the exceptional point. We call it ``hybrid point'' due to the anisotropy in the dispersion. We leave a systematical study of band degeneracies resulting from merging two exceptional points \cite{Keck2003, Kirillov2005} in Supplemental Material Sec.~VI. The remaining special case $ m=\kappa=0 $ hosts a ring of exceptional point at $ k=|\delta| $ \cite{Zhen2015}. This ``exceptional ring'' is present due to the rotational symmetry at $ \kappa=0 $, hence is generally unstable in two dimensions. 
As $ \delta $ tends to zero, the ring shrinks to a Dirac point. Only then do we recover the Hermitian topological phase transition point. 

In summary, the most general scenario of non-Hermitian topological transition is through ``hybrid point --- exceptional point pair --- hybrid point'', instead of the Dirac point in the Hermitian case.

As already been discussed earlier in this paper, the square root singularity in Eq.~\eqref{eq:disp} leads to the pair switching of eigenvalues/eigenstates around an exceptional point. This can be characterized by the half-integer quantized topological invariant $ \nu_\Gamma $ defined in Eq.~\eqref{eq:nu}, where $\Gamma$ encloses a single exceptional point. It follows from Eq.~\eqref{eq:disp} that $\nu_\Gamma=\pm 1/2 $ whose sign is determined by the sign of ${\rm Im} (c_y/c_x)$. Therefore, exceptional points are characterized by topological charges $\pm 1/2$.

We note that in Ref.~\cite{Leykam2017} there is a similar formula characterizing the topology of the exceptional point, which can be seen as a special case of Eq.~\eqref{eq:nu}, with the spectrum being symmetric with respect to $ E=0 $, i.e., $ a=0 $ in Eq.~\eqref{eq:Hep}. 
In Ref.~\cite{Xu2017}, the loop topology of exceptional points is characterized by the integral of the Berry phase when it is encircled twice. This can be seen as a special case of Eq.~\eqref{eq:nu} when the Hamiltonian is complex symmetric or of size $ 2\times 2 $. In general, this phase is a path-dependent geometric phase and is thus not quantized \cite{Berry1984,Garrison1988,Mailybaev2005}. 

Extension of non-Hermitian topological band theory to higher dimensions, different symmetry classes and its applications to a wide range of physical systems will be presented in forthcoming works.

\textit{Acknowledgment:} This work is supported by DOE Office of Basic Energy Sciences, Division of Materials Sciences and Engineering under Award DE-SC0010526. LF is partly supported by the David and Lucile Packard Foundation. 
HS is supported by MIT Alumni Fellowship Fund For Physics.
BZ is partially supported by the United States---Israel Binational Science Foundation (BSF) under award No. 2013508 and the Army Research Office through the Institute for Soldier Nanotechnologies under Contract No. W911NF-13-D-0001.

\bibliography{nonHermitian_Ref}

\widetext
\clearpage


\section*{Supplementary material}

# Supplemental Material for "Topological Band Theory for Non-Hermitian Hamiltonians"

Huitao Shen,[1] Bo Zhen,[1, 2] and Liang Fu[1]

[1]*Department of Physics, Massachusetts Institute of Technology, Cambridge, Massachusetts 02139, USA*
[2]*Department of Physics and Astronomy, University of Pennsylvania, Philadelphia, Pennsylvania 19104, USA*

## I. PROOF OF NON-ORTHOGONALITY OF LEFT AND RIGHT EIGENSTATES

Suppose the eigenvalues of $H$ are non-degenerate. This means the Jordan normal form of $H$ is diagonal: $H = PJP^{-1}$, where $P$ is an invertible matrix and $J$ is diagonal. First rewrite it as $HP = PJ$. Then the $n$-th column of $P$ is a right eigenstate of $H$, corresponding to the eigenvalue $E_n$, denote as $|\psi_n^R\rangle$. Then rewrite $HP = PJ$ as $H^\dagger (P^{-1})^\dagger = (P^{-1})^\dagger J^*$. Now the $m$-th column of $(P^{-1})^\dagger$ is a left eigenstate of $H$, corresponding to the eigenvalue $E_m^*$, denoted as $|\psi_m^L\rangle$. Finally, we notice that $P^{-1}P = I$. This means $\langle\psi_m^L|\psi_n^R\rangle = \delta_{mn}$ under this normalization condition. Under a general normalization condition $\langle\psi_n^L|\psi_n^R\rangle \neq 0$, i.e., the left and right eigenstates corresponding to the same eigenvalue are non-orthogonal.

## II. PROOF OF CHERN NUMBER EQUALITIES

Since the proof is for any given band, we suppress the band index $n$ in the following. Our proof relies on the fact the Chern number is an obstruction to a global gauge of the wavefunction in the whole BZ [1], and the gauges of the left and right eigenstates are "locked" to each other because $\langle\psi_n^L|\psi_n^R\rangle \neq 0$.

We first prove $N^{LR} = N^{RR}$. It is convenient to introduce the Berry connection for a given band:

$$\mathbf{A}^{\alpha\beta}(\mathbf{k}) = \langle \psi^\alpha(\mathbf{k})|\nabla_\mathbf{k} \psi^\beta(\mathbf{k})\rangle. \tag{1}$$

Here $\alpha, \beta = L/R$.

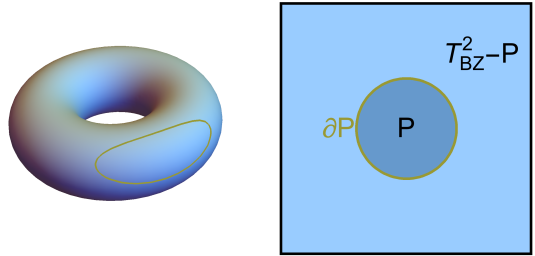

FIG. 1. (Color online) The schematic diagram of the patch $R$ in the 2D BZ.

Consider a patch $P$ in the BZ, whose boundary $\partial P$ has circumference $L$. A schematic diagram is shown in Fig. 1. We first choose a local gauge I in $P$ for the right eigenstate, denote as $|\psi_I^R\rangle$. We then choose another gauge II in $T_{BZ}^2 - P$, denote as $|\psi_{II}^R\rangle$. (We can always divide the BZ into more patches and follow the similar arguments below if two patches are not enough for well-defined local gauges. ) For the case $\alpha = \beta = R$, the two gauges are related by the gauge transformation

$$|\psi_{II}^R(\mathbf{k})\rangle = e^{if(\mathbf{k})} |\psi_I^R(\mathbf{k})\rangle \tag{2}$$

on the boundary $\partial R$. For the case $\alpha = L$ and $\beta = R$, on the boundary

$$\begin{aligned}|\psi_{II}^R(\mathbf{k})\rangle &= r(\mathbf{k})e^{if(\mathbf{k})} |\psi_I^R(\mathbf{k})\rangle, \\ |\psi_{II}^L(\mathbf{k})\rangle &= \frac{1}{r(\mathbf{k})} e^{if(\mathbf{k})} |\psi_I^L(\mathbf{k})\rangle.\end{aligned} \tag{3}$$

Both $r(\mathbf{k})$ and $f(\mathbf{k})$ are continuous real functions. Eq. (2) and Eq. (3) satisfy $\langle\psi^{\mathrm{R}}(\mathbf{k})|\psi^{\mathrm{R}}(\mathbf{k})\rangle = 1$ and $\langle\psi^{\mathrm{L}}(\mathbf{k})|\psi^{\mathrm{R}}(\mathbf{k})\rangle = 1$ respectively. Straightforward calculation gives the transformation law of the Berry connection

$$\mathbf{A}_{\mathrm{II}}^{\mathrm{RR}}(\mathbf{k}) = \mathbf{A}_{\mathrm{I}}^{\mathrm{RR}}(\mathbf{k}) + i\nabla_{\mathbf{k}}f(\mathbf{k}) \tag{4}$$

$$\mathbf{A}_{\mathrm{II}}^{\mathrm{LR}}(\mathbf{k}) = \mathbf{A}_{\mathrm{I}}^{\mathrm{LR}}(\mathbf{k}) + i\nabla_{\mathbf{k}}f(\mathbf{k}) + \frac{\nabla_{\mathbf{k}}r(\mathbf{k})}{r(\mathbf{k})}. \tag{5}$$

Berry curvature can be seen as the curl of the Berry connection: $\epsilon_{ij}B_{ij}^{\alpha\beta} = [\nabla\times\mathbf{A}]_z$, which is gauge independent due to $\nabla\times(\nabla f) = 0$. Stokes theorem implies

$$\begin{aligned}\int_{\mathrm{BZ}} \epsilon_{ij}B_{ij}^{\alpha\mathrm{R}}d^2\mathbf{k} &= \oint_{\partial R} (\mathbf{A}_{\mathrm{I}}^{\alpha\mathrm{R}} - \mathbf{A}_{\mathrm{II}}^{\alpha\mathrm{R}})\cdot d\mathbf{l} \\ &= i\oint_{\partial R}\nabla_{\mathbf{k}}f(\mathbf{k})\cdot d\mathbf{l} = i(f(L) - f(0)),\end{aligned} \tag{6}$$

where $\alpha = \mathrm{L/R}$. Notice that this number is quantized, pure imaginary, and is independent of the choice of $\alpha$. There is no dependence of $r(\mathbf{k})$ because

$$\oint_{\partial R}\frac{\nabla_{\mathbf{k}}r(\mathbf{k})}{r(\mathbf{k})}\cdot d\mathbf{l} = \oint_{\partial R}\nabla_{\mathbf{k}}\log r(\mathbf{k})\cdot d\mathbf{l} = \log\frac{r(L)}{r(0)} = 0. \tag{7}$$

Hence we have proved $N^{\mathrm{LR}} = N^{\mathrm{RR}}$. Physically, this is because the normalization condition $\langle\psi^\alpha(\mathbf{k})|\psi^{\mathrm{R}}(\mathbf{k})\rangle = 1$ fixes the gauge of $|\psi^\alpha(\mathbf{k})\rangle$, and hence the Chern number is solely determined by the gauge of the right eigenstate.

With the same spirit, we can prove $N^{\mathrm{RL}} = N^{\mathrm{LL}}$. Notice $B^{\mathrm{LR}}(\mathbf{k}) = -(B^{\mathrm{RL}}(\mathbf{k}))^\dagger$. Since the integral of the Berry curvature is purely imaginary according to Eq. (6), we have $N^{\mathrm{LR}} = N^{\mathrm{RL}}$. Combine all the equalities, we finally finished the proof $N^{\mathrm{RR}} = N^{\mathrm{LR}} = N^{\mathrm{RL}} = N^{\mathrm{LL}}$.

We note that although the Chern numbers are the same, the gauge invariant Berry curvatures are locally different. In fact, the left-right Berry curvature is a complex number while the left-left and right-right Berry curvatures are purely imaginary.

We have verified all the results in this section numerically using the generalized Dirac model Eq. (4) in the main text.

### III. VANISHING OF THE CHERN NUMBERS

Since we have proved $N^{\mathrm{RR}} = N^{\mathrm{LR}} = N^{\mathrm{RL}} = N^{\mathrm{LL}}$, we only consider $N^{\mathrm{LR}}$ in the following. According to Eq. (3) in the main text, the Chern numbers can vanish because

(a) The Berry curvature vanishes at all $\mathbf{k}$, i.e.,

$$\langle\partial_{k_x}\psi^{\mathrm{L}}(\mathbf{k})|\partial_{k_y}\psi^{\mathrm{R}}(\mathbf{k})\rangle = \langle\partial_{k_y}\psi^{\mathrm{L}}(\mathbf{k})|\partial_{k_x}\psi^{\mathrm{R}}(\mathbf{k})\rangle, \tag{8}$$

meaning $|\psi^{\mathrm{L}}(\mathbf{k})\rangle = |\psi^{\mathrm{R}}(\mathbf{k})\rangle^*$, which further indicates $H(\mathbf{k}) = H(\mathbf{k})^T$. This is equivalent to a complex symmetric Hamiltonian. For $2\times 2$ Hamiltonians $H = \mathbf{b}\cdot\boldsymbol{\sigma}$, this means $b_y = 0$. Only two Pauli matrices are allowed.

(b) The Berry curvature cancels in the whole BZ, i.e.

$$\begin{aligned}&\langle\partial_{k_x}\psi^{\mathrm{L}}(\mathbf{k})|\partial_{k_y}\psi^{\mathrm{R}}(\mathbf{k})\rangle - \langle\partial_{k_y}\psi^{\mathrm{L}}(\mathbf{k})|\partial_{k_x}\psi^{\mathrm{R}}(\mathbf{k})\rangle \\ &= \langle\partial_{k_y}\psi^{\mathrm{L}}(-\mathbf{k})|\partial_{k_x}\psi^{\mathrm{R}}(-\mathbf{k})\rangle - \langle\partial_{k_x}\psi^{\mathrm{L}}(-\mathbf{k})|\partial_{k_y}\psi^{\mathrm{R}}(-\mathbf{k})\rangle.\end{aligned} \tag{9}$$

This indicates $|\psi^{\mathrm{L}}(\mathbf{k})\rangle = |\psi^{\mathrm{R}}(-\mathbf{k})\rangle^*$, further indicating $H(\mathbf{k}) = H(-\mathbf{k})^T$.

The relation of these conditions with the symmetry of the Hamiltonian will be discussed in a separate paper.

### IV. BULK-EDGE CORRESPONDENCE IN 2D

In this section, we first discuss the numerical solution of the domain wall problem (Eq. (5) in the main text) in detail. Then we show the numerical result of a similar problem on the square lattice to strengthen the discussion.

In order for the ansatz Eq. (6) in the main text to satisfy Eq. (5) in the main text, the following equations and conditions should be satisfied:

(i) Wavefunctions at both domains have the same energy:

$$E(k) \equiv \sqrt{(m_1 + i\delta_1)^2 + (k + i\kappa_{1,y})^2 - (1/\lambda_+ - \kappa_{1,x})^2} \\ = \sqrt{(m_2 + i\delta_2)^2 + (k + i\kappa_{2,y})^2 - (1/\lambda_- - \kappa_{2,x})^2}, \quad (10)$$

(ii) Continuity of the wavefunction at $x = 0$:

$$\frac{(m_1 + i\delta_1) \pm E(k)}{(k + i\kappa_{1,y}) - (1/\lambda_+ - \kappa_{1,x})} = \frac{(m_2 + i\delta_2) \pm E(k)}{(k + i\kappa_{2,y}) - (1/\lambda_- - \kappa_{2,x})}, \quad (11)$$

(iii) Localized edge state

$$\text{Re}(1/\lambda_+) > 0, \ \text{Re}(1/\lambda_-) < 0. \quad (12)$$

$E(k)$ is generally complex and this makes an analytical solution hard to obtain. The imaginary part of $E(k)$ depends on $\kappa_y$ and $\delta$, and vanishes only under two circumstances: (a) When $\kappa_y = 0$. The reason is that $\kappa_x$ only gives the mode propagating perpendicular to the domain wall a dissipation, since the boundary is along the $y$ direction; (b) When the gain and loss provided by the two domains cancels delicately. For example, the imaginary part $\text{Im}(E(k=0))$ as a function of $\kappa_{1,y}$ and $\kappa_{2,y}$ is plotted in Fig. 2. It vanishes only when $\kappa_{1,y} + \kappa_{2,y} = 0$.

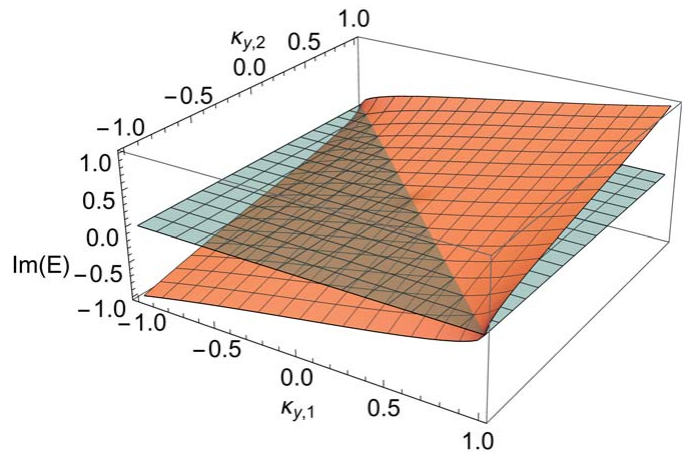

FIG. 2. (Color online) The imaginary part of the edge state energy $\text{Im}(E)$ at $k = 0$ as a function of $\kappa_{1,y}$ and $\kappa_{2,y}$. The transparent blue surface is the $\text{Im}(E) = 0$ equal energy surface. Here $m_1 = -m_2 < 0$ and $\kappa_{1,x} = \kappa_{2,x} = \delta_1 = \delta_2 = 0$. The energy unit is $|m_1| = |m_2|$.

To strengthen the discussion of the topological phase transition in the continuous Dirac model, here we consider a lattice model on a square lattice. The bulk Hamiltonian is given by

$$H(\mathbf{k}) = (t \sin k_x + i\kappa_x)\sigma_x + (t \sin k_y + i\kappa_y)\sigma_y \\ + (\cos k_x + \cos k_y + m + i\delta)\sigma_z. \quad (13)$$

In the following, $t = 1$ is the energy unit. In the Hermitian limit $\kappa_x = \kappa_y = 0$, this model hosts a topological phase transition. $|m| < 2$ corresponds to the topologically nontrivial phase and $|m| > 2$ corresponds to the topological trivial phase. There are band degeneracies at $|m| = 2$ where four Dirac fermions are located on $\mathbf{k} = (0,0), (0,\pi), (\pi,0)$ and $(\pi,\pi)$.

In order to see the effect of edge states, we numerically compute the complex spectrum of this model under a cylinder geometry. The lattice is periodic along $y$ direction, and has a finite number of $n$ sites along $x$ direction. In Fig. 3 we present the complex spectrum of a topologically nontrivial system with $n = 40$. In both of the panels, the bulk spectrum is separable. We can see clearly that two chiral edge states within the complex-energy "gap" connect the bulk bands. In the upper panel, the edge states do not dissipate. In the lower panel, the edge states have a constant imaginary energy due to the nonzero $\kappa_y$. These results are consistent with those of the domain wall problem perfectly discussed in the main text.

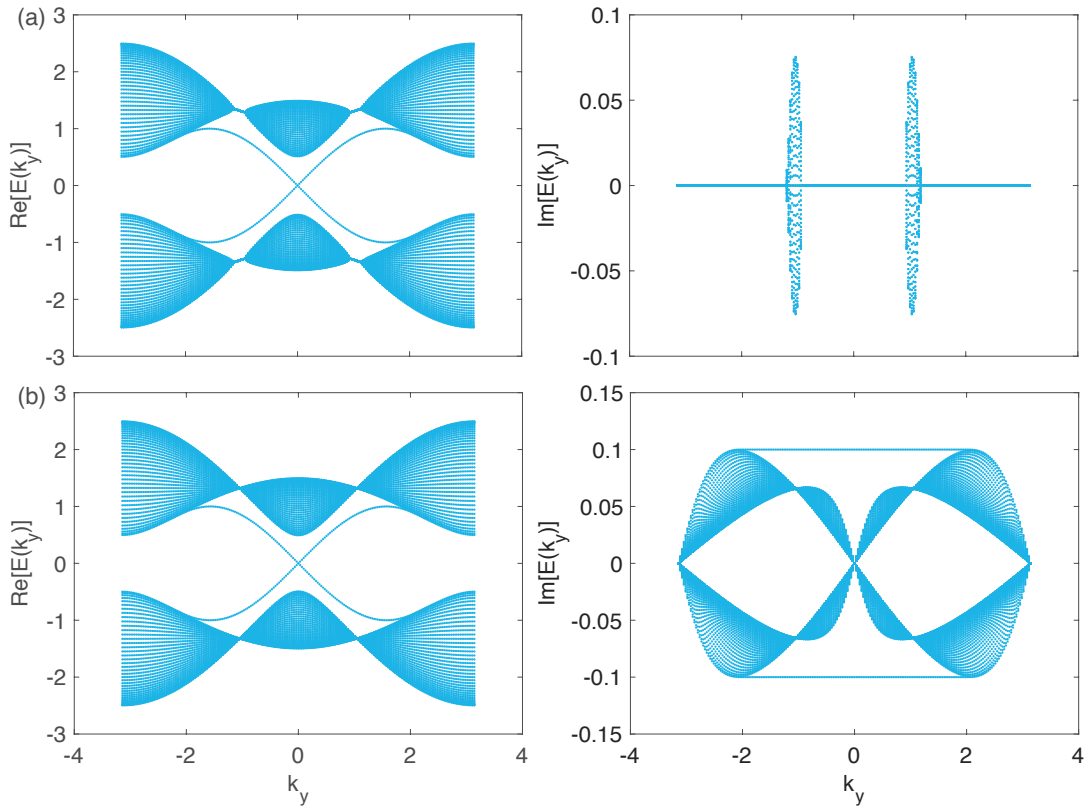

FIG. 3. (Color online) The complex energy spectrum of Eq. (13). The energy unit is the hopping strength $t$. The number of sites along x-direction is $n = 40$. The system is in the topologically nontrivial phase $m = -0.5, \delta = 0$. (a) $\kappa_x = 0.1$, $\kappa_y = 0$; (b) $\kappa_x = 0$, $\kappa_y = 0.1$.

## V. NON-HERMITIAN LEVEL DEGENERACY REQUIRES TUNING 2 PARAMETERS

In this section, we review the well-established result that finding a level degeneracy in non-Hermitian Hamiltonians generically requires tuning 2 parameters [2].

It suffices to consider an $2 \times 2$ effective Hamiltonian $H$ near the degeneracy point, obtained by projecting the generic Hamiltonian $\tilde{H}$ to the subspace of these two levels. The most general form of $H$ is

$$H = aI + (\mathbf{b}_0 + i\mathbf{b}_1) \cdot \boldsymbol{\sigma}, \tag{14}$$

where $I$ is the identity matrix, $\boldsymbol{\sigma} \equiv (\sigma_x, \sigma_y, \sigma_z)$ denotes Pauli matrices, parameter $a$ is complex, and $\mathbf{b}_i = (b_{i,x}, b_{i,y}, b_{i,z})$ with $i = 0, 1$ are real. The spectrum of $H$ is

$$E_\pm = a \pm \sqrt{\mathbf{b}_0 \cdot \mathbf{b}_0 - \mathbf{b}_1 \cdot \mathbf{b}_1 + 2i\mathbf{b}_0 \cdot \mathbf{b}_1}. \tag{15}$$

Degeneracy of eigenvalues ($E_+ = E_-$) occurs when

$$\mathbf{b}_0 \cdot \mathbf{b}_0 = \mathbf{b}_1 \cdot \mathbf{b}_1, \text{ and } \mathbf{b}_0 \cdot \mathbf{b}_1 = 0. \tag{16}$$

The two equations above, which are the necessary and sufficient condition for level degeneracy, can be generically satisfied by tuning 2 real parameters in the Hamiltonian. In the Hermitian case when $\mathbf{b}_1 = 0$, these two equations reduce to $\mathbf{b}_0 = 0$, requiring to tune all 3 components of $\mathbf{b}_0$.

At a degeneracy point, Eq. (16) implies the two vectors $\mathbf{b}_0$ and $\mathbf{b}_1$ are orthogonal and of equal length. Hence, without loss of generality we choose $\mathbf{b}_0 = b(1, 0, 0)$ and $\mathbf{b}_1 = b(0, 1, 0)$. The resulting non-Hermitian Hamiltonian is

$$H_0 = aI + b\sigma_+. \tag{17}$$

This can be considered as the most general Hamiltonian at a non-Hermitian degeneracy. Importantly, for $b \neq 0$, the above $2 \times 2$ matrix has only one eigenstate $(1, 0)$. Such a non-Hermitian Hamiltonian is called "defective", Only for the special case $b = 0$, $H_0$ is proportional to identity and has two degenerate and linearly-independent eigenstates. This result shows that the generic degeneracy points of non-Hermitian Hamiltonians are defective.

## VI. MERGING TWO EXCEPTIONAL POINTS

In this section, we systematically study all four different scenarios when two exceptional points are moved together. Recall (Eq. (11) and (12) in the main text) that the most general Hamiltonian near the exceptional point could be written as $H(\mathbf{k}) = aI + \epsilon\sigma_+ + \sum_{i,j} k_i c_{ij}\sigma_j$, with the dispersion

$$E_\pm(\mathbf{k}) = a \pm \sqrt{c_x k_x + c_y k_y}, \tag{18}$$

where $c_x = 2\epsilon(c_{xx} + ic_{xy})$ and $c_y = 2\epsilon(c_{yx} + ic_{yy})$.

The first two scenarios involve moving together two exceptional points with opposite charge, which we have already encountered when discussing the topological phase transitions.

(a) It is possible that $c_x = c_y = 0$ in Eq. (18) and there is no phase accumulation in the quadratic terms. This is actually the case when two exceptional points merge into the familiar Dirac point in the Hermitian Hamiltonians. The Hamiltonian

$$H = (k_x + i\delta)\sigma_x + k_y\sigma_y, \tag{19}$$

hosts two exceptional points at $\mathbf{k} = (0, \pm\delta)$ with vorticity $\pm 1/2$ respectively. As $\delta \to 0$, the exceptional points annihilate into the Dirac point, where there are neither interchange of the eigenvalues, nor the Hamiltonian is defective.

(b) There is another case when $c_x$ and $c_y$ in Eq. (18) have the same argument $\text{Im}(c_y/c_x) = 0$ so that there is no net phase accumulation when the degeneracy is encircled. With a proper rotation of the momentum to $k'_x$ and $k'_y$, the dispersion must be proportional to $\sqrt{k'_x}$ at one direction and to $k'_y$ in another direction. Based on the dispersion profile, we call this degeneracy "hybrid point". We demonstrate hybrid point with the Hamiltonian

$$H = (k_x + i\delta)\sigma_x + k_y\sigma_y + m\sigma_z, \quad |\delta| \geq |m| > 0. \tag{20}$$

It has two exceptional points at $\mathbf{k} = (0, \pm\sqrt{\delta^2 - m^2})$ with opposite charges. When $|\delta| = |m|$, the two exceptional points merge at $\mathbf{k} = 0$ and the spectrum becomes $E_\pm(\mathbf{k}) = \pm\sqrt{k_x^2 + k_y^2 + 2ik_x m}$ (Fig. 4). It is obvious there is no interchange of the eigenvalues, although the Hamiltonian at this point is still defective.

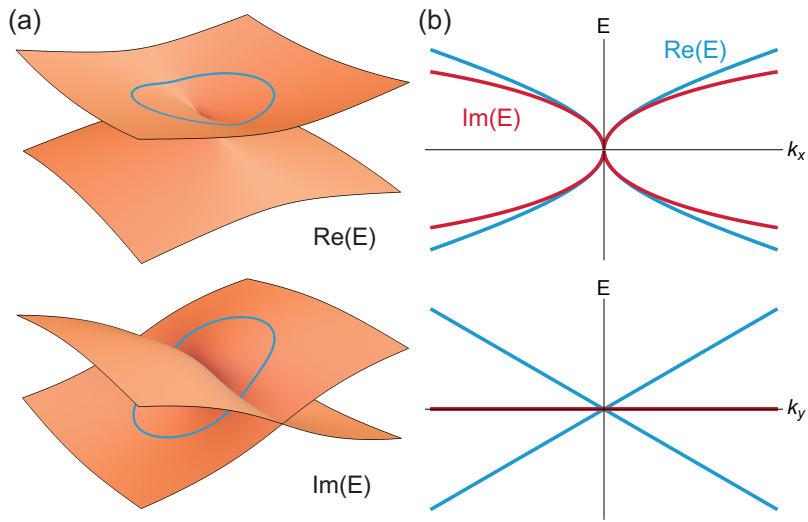

FIG. 4. (Color online) (a) The dispersion near a hybrid point. The Hamiltonian is Eq. (20) for $m = \delta = 1$. The blue loop is $k = 1$. Clearly there is no eigenvalue switching. (b) Same dispersion along $k_y = 0$ and $k_x = 0$, when the dispersion is proportional to $\sqrt{|k_x|}$ and $|k_y|$ respectively. This anisotropy is why the degeneracy is called "hybrid point".

Moving together two exceptional points with the same charge gives the remaining two scenarios.

(c) The degeneracy is a "double exceptional point" with vorticity $\pm 1$, on which the Hamiltonian is defective. This can be seen from

$$H = (k_x + isk_y)\sigma_z + (\delta + \lambda)\sigma_x + i\delta\sigma_y, \quad \delta \neq 0. \tag{21}$$

Here $s = \pm 1$. There are two vorticity $s/2$ exceptional points at $\mathbf{k} = (0, \pm\sqrt{\lambda(\lambda + 2\delta)})$. As $\lambda \to 0$, $\mathbf{k} = 0$ is a double exceptional point.

(d) The Hamiltonian is not necessarily defective when the vorticity is an integer. Consider the previous Hamiltonian Eq. (21) with $\delta = 0$. The Hamiltonian at $\lambda \to 0$ is now just a zero matrix, which is of course not defective. Although there is no interchanging of the eigenstates, the eigenvalues have a integer phase winding like a vortex. We thus name this point "vortex point".

TABLE I. Four types of degeneracy in non-Hermitian Hamiltonians with their properties.

| Degeneracy | Vorticity | Defectiveness |
|---|---|---|
| Exceptional point | Half-integer | Defective |
| Hybrid point | Zero | Defective |
| Dirac point | Zero | Non-defective |
| Vortex point | Nonzero integer | Non-defective |

All the four types of degeneracy are summarized in the Table. I. The non-zero winding of the eigenvalues is not necessarily related to the defectiveness of the Hamiltonian. Except the exceptional point, all three other types of degeneracy need fine-tuning parameters and any perturbation will make them into two exceptional points in 2D.


\end{document}